\documentclass[fleqn,twoside]{article}
\usepackage{amsmath}
\usepackage{espcrc2}
\usepackage{amssymb}
\usepackage{psfrag}
\usepackage{graphicx}
\usepackage{latexsym,amsfonts}
\begin{document}

\title{Rates of K--shell Electron Capture Decays of ${^{180}}{\rm Re}$
and ${^{142}}{\rm Pm}$ Atoms} \author{A. N. Ivanov$^a$\thanks{\mbox{\it E-mail }:
ivanov@kph.tuwien.ac.at}, P. Kienle$^{b,c}$\thanks{\mbox{\it E-mail }:
Paul.Kienle@ph.tum.de}, M. Pitschmann$^a$\thanks{E--mail:
pitschmann@kph.tuwien.ac.at},\\ \addressmark{$^a$Atominstitut der
\"Osterreichischen Universit\"aten, Technische Universit\"at Wien,
Wiedner Hauptstra\ss e 8-10, A-1040 Wien, Austria, \\ $^b$Stefan Meyer
Institut f\"ur subatomare Physik, \"Osterreichische Akademie der
Wissenschaften, Boltzmanngasse 3, A-1090, Wien, Austria},\\
$^c$Excellence Cluster Universe Technische Universit\"at M\"unchen,
D-85748 Garching, Germany}

\date{\today}

\begin{abstract}
  We propose a theoretical analysis of the experimental data on the
  time behaviour K--shell electron capture $(EC)$ decays of atoms
  ${^{180}}{\rm Re}$ and ${^{142}}{\rm Pm}$ in solid targets, obtained
  recently by F\"astermann {\it et al.}, Phys. Lett. B {\bf 672}, 227
  (2009) and Vetter {\it et al.}, Phys. Lett. B {\bf 670}, 149 (2008).
  We show that the absence of the time modulation in these data rules
  out the explanation of the ``GSI Oscillations'' (Yu. A. Litvinov
  {\it et al.}, Phys. Lett. B {\bf 664}, 162 (2008)) by means of two
  closely spaced ground mass--eigenstates of mother nuclei.  PACS:
  12.15.Ff, 13.15.+g, 23.40.Bw, 26.65.+t
\end{abstract}

\maketitle

\subsection*{Introduction}

Measurements of the K--shell electron capture ($EC$) and positron
($\beta^+$) decay rates of the H--like heavy ions ${^{142}}{\rm
Pm}^{60+}$, ${^{140}}{\rm Pr}^{58+}$, and ${^{122}}{\rm I}^{52+}$ with
one electron on their K--shells have been recently carried out in the
Experimental Storage Ring (ESR) at GSI in Darmstadt
\cite{GSI2}--\cite{GSI5}. The measurements of the rates of the number
of daughter ions ${^{142}}{\rm Nd}^{60+}$, ${^{140}}{\rm Ce}^{58+}$
and ${^{122}}{\rm Te}^{52+}$ showed a time modulation of exponential
decays with periods $T_{EC} = 7.10(22)\,{\rm s}, 7.06(8)\,{\rm s}$ and
$6.11(3)\,{\rm s}$ and modulation amplitudes $a^{EC}_d = 0.23(4),
0.18(3)$ and $0.22(2)$ for ${^{142}}{\rm Pm}^{60+}$, ${^{140}}{\rm
Pr}^{58+}$ and ${^{122}}{\rm I}^{52+}$, respectively.

Since the rates of the number $N^{EC}_d(t)$ of daughter ions are
defined by
\begin{eqnarray}\label{label1}
  \frac{dN^{EC}_d(t)}{dt} = \lambda^{EC}_d(t)\, N_m(t),
\end{eqnarray}
where $\lambda^{EC}_d(t)$ is related to the $EC$--decay rate and
$N_m(t)$ is the number of mother ions at time $t$ after the mother
ions are produced, the time--modulation of $dN^{EC}_d(t)/dt$ can be
described in terms of a periodic time--dependence of the $EC$--decay
rate $\lambda^{EC}_d(t)$ \cite{GSI2}--\cite{GSI5}
\begin{eqnarray}\label{label2}
  \lambda^{EC}_d(t) = \lambda_{EC}\,(1 + a^{EC}_d \cos(\omega_{EC}t +
\phi_{EC})),
\end{eqnarray}
where $\lambda_{EC}$ is the $EC$--decay constant, $a^{EC}_d$, $T_{EC}
= 2\pi/\omega_{EC}$ and $\phi_{EC}$ are the amplitude, period and
phase of the time--dependent term \cite{GSI2}. Furthermore it was
shown that the $\beta^+$--decay rate of ${^{142}}{\rm Pm}^{60+}$,
measured simultaneously with its modulated $EC$--decay rate, is not
modulated with an amplitude upper limit $a_{\beta^+} < 0.03$
\cite{GSI3}--\cite{GSI5}.

The important property of the periods of the time modulation of the
$EC$--decay rates is their proportionality to the mass number $A$ of
the nucleus of the mother H--like heavy ion. Indeed, the periods
$T_{EC}$ can be described well by the phenomenological formula $T_{EC}
= A/20\,{\rm s}$.  The proportionality of the periods of the
$EC$--decay rates to the mass number $A$ of the mother nuclei and the
absence of the time modulation of the $\beta^+$--decay branch
\cite{GSI3}--\cite{GSI5} can be explained, assuming the interference
of neutrino mass--eigenstates caused by a coherent superposition of
mono--chromatic neutrino mass--eigenstates with electron lepton charge
\cite{Ivanov2,Ivanov4}. Indeed, nowadays the existence of massive
neutrinos, neutrino--flavour mixing and neutrino oscillations is well
established experimentally and elaborated theoretically \cite{PDG08}.
The observation of the interference of massive neutrino
mass--eigenstates in the $EC$--decay rates of the H--like heavy ions
sheds new light on the important properties of these states.

According to atomic quantum beat experiments \cite{QB}, the
explanation of the ``GSI oscillations'', proposed in
\cite{Ivanov2,Ivanov4}, bears similarity with quantum beats of atomic
transitions, when an excited atomic eigenstate decays into a coherent
state of two (or several) lower lying atomic eigenstates. In the case
of the $EC$--decay one deals with a transition from the initial state
$|m\rangle$ to the two--body final state $|d\,\nu_e\rangle$ with a
daughter ion $d$ in its stable ground state and an electron neutrino
$\nu_e$, which is a coherent superposition of two neutrino
mass--eigenstates with the energy difference equal to $\omega_{21} =
\Delta m^2_{21}/2 M_m $ related to $\omega_{EC}$ in the moving ion
system with a Lorentz factor $\gamma = 1.43$ as $\omega_{EC} =
\omega_{21}/\gamma$, where $\Delta m^2_{21} = m^2_2 - m^2_1$ is the
difference of squared neutrino masses $m_2$ and $m_1$ of
mass--eigenstates $\nu_2$ and $\nu_1$, respectively.

Another mechanism of the ``GSI oscillations'' has been proposed by
Giunti \cite{Giunti2,Giunti3} and Kienert {\it et al.}
\cite{Lindner}. The authors \cite{Giunti2}--\cite{Lindner} assume the
existence of two closely spaced ground mass--eigenstates of the
nucleus of the H--like heavy ion of unknown origin as the initial
state of the $EC$--decay and describe it by a coherent superposition
of the wave functions of two closely spaced ground mass--eigenstates
\begin{eqnarray}\label{label3}
|m\rangle = \cos\theta\,|m'\rangle + \sin\theta\,|m''\rangle,
\end{eqnarray}
where $|m'\rangle$ and $|m ''\rangle$ are two closely spaced ground
mass--eigenstates of the nucleus of the mother ion with masses $M_{m
'}$ and $M_{m ''}$, respectively, $\theta$ is a mixing angle. By
definition of eigenstates the mass--eigenstates of the nucleus of the
H--like heavy ion $|m'\rangle$ and $|m ''\rangle$ should be orthogonal
$\langle m'|m''\rangle = 0$.

Unlike our analysis \cite{Ivanov2,Ivanov4}, the authors
\cite{Giunti2}--\cite{Lindner} draw an analogy of the ``GSI
oscillations'' with quantum beats of atomic transitions \cite{QB},
when an atom, excited into a state of a coherent superposition of two
closely spaced energy eigenstates, decays into a lower lying energy
eigenstate. According to \cite{QB}, the intensity of radiation, caused
by a transition from such a coherent state into a lower energy
eigenstate, has a periodic time--dependent term with a period
inversely proportional to the energy--difference $\Delta E_{m'm''}$
between two closely spaced energy eigenstates. In the approach,
proposed in \cite{Giunti2}--\cite{Lindner}, the period of the time
modulation is equal to $T_{EC} = 2\pi\gamma/\Delta E_{m'm''}$, where
$\Delta E_{m'm''} = M_{m'} - M_{m''} = 8.38(9)\times 10^{-16}\,{\rm
eV}$ fixed for $T_{EC} = 7.06(8)\,{\rm s}$ and $\gamma = 1.43$, the
period of the time modulation of the $EC$--decay rate and the Lorentz
factor of the H--like ${^{140}}{\rm Pr}^{58+}$ ion \cite{GSI2}.

Such an explanation of the ``GSI oscillations'' is ruled out by the
experimental data on the time modulation of (i) the $EC$--decay rates
of the H--like ${^{122}}{\rm I}^{52+}$ ions, showing the time
modulation with the period $T_{EC} = 6.11(3)\,{\rm s}$ and obeying the
so--called $A$--scaling $T_{EC} = A/20\,{\rm s}$ of periods of the
time modulation of the $EC$--decay rates, where $A$ is the mass number
of mother nuclei \cite{GSI3}--\cite{GSI5}, and (ii) the positron
($\beta^+$) decay rates of the H--like ${^{142}}{\rm Pm}^{60+}$ ions,
showing no time modulation \cite{GSI3}--\cite{GSI5}. Indeed, as has
been shown in \cite{Ivanov5}, in case of the existence of two closely
spaced ground mass--eigenstates of the nucleus of the H--like mother
ion the $\beta^+$--decay rates of the H--like heavy ions should be
modulated with a period $T_{\beta^+}$ equal to the period of the time
modulation of the $EC$--decay rate, i.e. $T_{\beta^+} = T_{EC} =
7\,{\rm s}$. 

In this letter we give one more reason for the exclusion of the
explanation of the ``GSI oscillations'' by means of two closely spaced
ground mass--eigenstates of the nuclei of mother ions, proposed in
\cite{Giunti2}--\cite{Lindner}.
\subsection*{Experimental analysis of $EC$--decay rate of ${^{180}}{\rm Re}$ atoms} 
Recently \cite{GSI6} the experimental data on the time modulation of
the $EC$--decay rate of ${^{180}}{\rm Re}$ atoms with
quantum numbers $I^{\pi} = 1^-$, where $I$ is the nuclear spin, have
been declared as a new test for theoretical approaches for the
explanation of the ``GSI oscillations''. 

In \cite{GSI6} atoms ${^{180}}{\rm Re}$, produced in the reaction
${^{181}}{\rm Ta}({^3}{\rm He},4n)$ in the $50\,{\rm mg/cm^2}$
tantalum foil with a $33\,{\rm MeV}$ ${^3}{\rm He}$ beam from the
Munich MLL tandem accelerator \cite{GSI6}, are unstable under the
$EC$--decay ${^{180}}{\rm Re}_{I^{\pi} = 1^-} \to {^{180}}{\rm
W}_{I^{\pi} = 2^-} + \nu_e$ with nuclei ${^{180}}{\rm W}_{I^{\pi} =
2^-}$ in the excited state, which then decay ${^{180}}{\rm W}_{I^{\pi}
= 2^-} \to {^{180}}{\rm W}_{I^{\pi} = 2^+} + \gamma$ with emission of
photons \cite{ND96}.  According to the experimental $A$--scaling of
the periods of time modulation \cite{GSI3,GSI4}, the rate of the
number of daughter atoms or the radiated intensity should show a time
modulation with a period $T_{EC} = A /20\gamma = 6.3\,{\rm
s}$. However, the experimental time spectrum of the $EC$--decays of
${^{180}}{\rm Re}$ shows no time modulation (see Fig.\,1) \cite{GSI6}.
\begin{figure}[t]
\includegraphics[width = 0.70\linewidth]{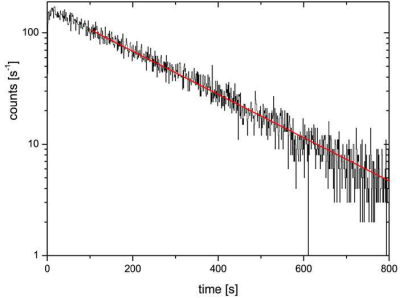}
\caption{The experimental time spectrum of the $EC$--decay of the
${^{180}}{\rm Re}$ ion measured in \cite{GSI6}. The time axis is
binned with $1\,{\rm s/\rm bin}$. The result of a fit with purely
exponential decay is also shown \cite{GSI6}.}
\end{figure}
\subsection*{Experimental analysis of $EC$--decay rate of ${^{142}}{\rm Pm}$ atoms}
In Ref.\cite{PV08} the authors have investigated a time modulation of
the $EC$--decay rate of atoms ${^{142}}{\rm Pm}$ with quantum numbers
$I^{\pi} = 1^+$ in the transitions ${^{142}}{\rm Pm}_{I^{\pi} = 1^+}
\to {^{140}}{\rm Nd^*}_{I^{\pi} = 0^+} + \nu_e \to {^{140}}{\rm
Nd}_{I^{\pi} = 0^+} + \gamma + \nu_e$, where the K--holes in the atoms
${^{140}}{\rm Nd^*}$ are occupied by the electrons from the
$L_{II}$--shell with the emission of photons. The atoms ${^{142}}{\rm
Pm}$ were produced in the ${^{124}}{\rm Sn} ({^{23}}{\rm Na},
5n){^{142}}{\rm Pm}$ reaction at the Berkeley $88$--inch Cyclotron
with a bombarding time short compared to the periods of the time
modulation measured in \cite{GSI2}, i.e of order of $T_{EC}/1.43
\simeq 5\,{\rm s}$.  A time spectrum of $X$--rays from the transition
${^{142}}{\rm Nd^*}_{I^{\pi} = 0^+} \to {^{142}}{\rm Nd}_{I^{\pi} =
0^+} + \gamma$ has been measured.  As has been found the number of
counts can be fitted well by exponential decay curve and shows no time
modulation (see Fig.\,2).
\begin{figure}[t]
\includegraphics[width = 0.80\linewidth]{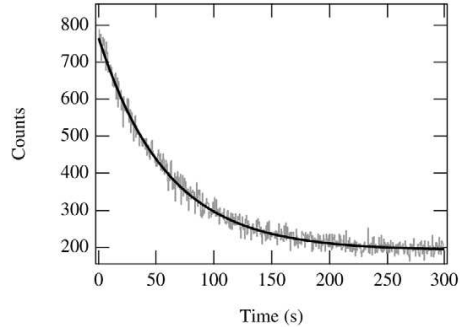}
\caption{Decay of ${^{142}}{\rm Nd}$ $K_{\alpha}$ $X$--rays. The
number of counts in a window surrounding the $K_{\alpha}$ peak is
plotted as a function of time after production of ${^{142}}{\rm
Pm}$. The best fit exponential decay curve is also shown \cite{PV08}.}
\end{figure}

There has been also found no time modulation for the $\beta^+$--decay
rates of atoms ${^{142}}{\rm Pm}$.  This agrees well with the
prediction given in \cite{Ivanov4} for the H--like heavy ions
${^{142}}{\rm Pm}^{60+}$.

Below we give a theoretical analysis of the experimental data on the
$EC$--decays of ${^{180}}{\rm Re}$ and ${^{142}}{\rm Pm}$ atoms.
\subsection*{Transition amplitudes and rates}
For the analysis of the time modulation of the decay rates of the
${^{180}}{\rm Re}_{I^{\pi} = 1^-} \to {^{180}}{\rm W}_{I^{\pi} = 2^-}
+ \nu_e \to {^{180}}{\rm W}_{I^{\pi} = 2^+} + \gamma + \nu_e$ and
${^{142}}{\rm Pm}_{I^{\pi} = 1^+} \to {^{142}}{\rm Nd^*}_{I^{\pi} =
0^+} + \nu_e \to {^{142}}{\rm Nd}_{I^{\pi} = 0^+} + \gamma + \nu_e$
decays we follow time--dependent perturbation theory \cite{QM} and the
procedure, proposed in \cite{WG96}, for the theoretical analysis of
the chains of the transitions.  

The transitions are caused by the Hamilton operators $H_W(t'')$ and
$H_{\rm elm}(t')$ of weak and electromagnetic interactions
\begin{eqnarray}\label{label4}
  \hspace{-0.3in}&&H_W(t'') = \nonumber\\
\hspace{-0.3in}&&= \frac{G_F}{\sqrt{2}}V_{ud}\int d^3x''\,
  [\bar{\psi}_n(x'')\gamma^{\mu}(1 - g_A\gamma^5)
  \psi_p(x'')]\nonumber\\
\hspace{-0.3in}&&\times [\bar{\psi}_{\nu_e}(x'') \gamma_{\mu}(1 -
  \gamma^5)\psi_{e^-}(x'')],\nonumber\\
\hspace{-0.3in}&&H_{\rm elm}(t') = e\int d^3x'\, \vec{J}(x')\cdot
\vec{A}(x')
\end{eqnarray}
with standard notation \cite{Ivanov1,AA94}, $\vec{J}(x')$ and
$\vec{A}(x')$ are the nuclear electromagnetic current and the
electromagnetic vector potential, respectively \cite{AA94}. 

The amplitudes of the decays under consideration we define as follows
\cite{WG96}
\begin{eqnarray}\label{label5}
\hspace{-0.3in}&&A(m \to d_+ \gamma\, \nu_e)(t) = -\int^t_{-\infty}dt'
\langle \gamma d_+|H_{\rm elm}(t')|d_-\rangle\nonumber\\
\hspace{-0.3in}&&\times\, \int^{t'}_{-\infty}dt''\langle \nu_e
d_-|H_W(t'')|m\rangle,
\end{eqnarray}
where $m$ is the mother atom ${^{180}}{\rm Re}$ or ${^{142}}{\rm Pm}$,
$d_-$ is the daughter atom ${^{180}}{\rm W}_{I^{\pi} = 2^-}$ or
${^{142}}{\rm Nd^*}_{I^{\pi} = 0^+}$, $\nu_e$ is the electron
neutrino, $ d_+$ is the daughter atom ${^{180}}{\rm W}_{I^{\pi} =
2^+}$ or ${^{142}}{\rm Nd}_{I^{\pi} = 0^+}$ and $\gamma$ is a photon.
For the calculation of the integrals over time we use the
$\varepsilon$--regularization \cite{Ivanov4}. 

The transition amplitudes $A(m \to d_+ \gamma\,\nu_e)(t)$ we calculate
below for (i) the initial state of the mother ion $m$, defined by a
coherent superposition of two closely spaced ground mass--eigenstates
$|m'\rangle$ and $|m''\rangle$, given by Eq.(\ref{label3}), and
massless electron neutrino $|\nu_e\rangle$, and (ii) the mother ion
$m$ in the one initial state and the electron neutrino, given by a
coherent superposition of neutrino mass--eigenstates $|\nu_j\rangle$
as $|\nu_e \rangle = \sum_j U^*_{ej}|\nu_j\rangle$, where $U^*_{ej}$
are matrix elements of the mixing matrix $U$ \cite{PDG08}. In the last
case the Hamilton operator of weak interactions is changed as
$H_W(t'') = \sum_j U_{ej}H^{(j)}_W(t'')$, where $H^{(j)}_W(t'')$ is
\cite{Ivanov2,Ivanov4}
\begin{eqnarray}\label{label6}
  \hspace{-0.3in}&&H^{(j)}_W(t'') = \nonumber\\
\hspace{-0.3in}&&= \frac{G_F}{\sqrt{2}}V_{ud}\int d^3x''\,
  [\bar{\psi}_n(x'')\gamma^{\mu}(1 - g_A\gamma^5)
  \psi_p(x'')]\nonumber\\
\hspace{-0.3in}&&\times [\bar{\psi}_{\nu_j}(x'') \gamma_{\mu}(1 -
  \gamma^5)\psi_{e^-}(x'')].
\end{eqnarray}
The rate $\lambda^{(\gamma)}_{EC}(t)$ of the $m \to d_+ + \gamma +
\nu_e$ transition is defined by
\begin{eqnarray}\label{label7}
\lambda^{(\gamma)}_{EC}(t) \propto \int\sum\frac{d^2}{dt^2}|A(m \to
d_+ \gamma \nu_e)(t)|^2\,d\rho_f,
\end{eqnarray}
where we sum over all polarisations of interacting particles and
$d\rho_f$ is the element of the phase volume of the final state $f =
d_+\gamma \,\nu_e$.

\subsubsection*{$EC$--decay from two 
closely spaced mass--eigenstates of mother nucleus ${^{180}}{\rm Re}$}

In the approach, based on the wave function of the mother atom given
by Eq.(\ref{label3}), for the integrand of the $EC$--decay rate of the
$m \to d_- + \nu_e \to d_+ + \gamma + \nu_e$ decay we get the
following expression
\begin{eqnarray}\label{label8}
  \hspace{-0.3in}&&\lim_{\varepsilon \to 0}\sum\frac{d^2}{dt^2}|A(m
\to d_+ \gamma \nu_e)(t)|^2 \propto\nonumber\\
 \hspace{-0.3in}&&\propto \delta(\omega + E_{d_+}(\vec{k}_+) -
E_{d_-}(\vec{k} +
\vec{k}_+))\nonumber\\\hspace{-0.3in}&&\times\,\delta(E_{\nu_e}(\vec{k}_{\nu_e})
+ E_{d_-}(\vec{k} + \vec{k}_+) -
M_m)\nonumber\\\hspace{-0.3in}&&\times\,\delta^{(3)}(\vec{k} +
\vec{k}_+ + \vec{k}_{\nu_e}) \nonumber\\
 \hspace{-0.3in}&&\times\, (1 + \sin 2 \theta\,\cos(\Delta
 E_{m'm''}t)).
\end{eqnarray}
Substituting Eq.(\ref{label8}) into Eq.(\ref{label7}) and integrating
over the phase volume we get
\begin{eqnarray}\label{label9}
  \lambda^{(\gamma)}_{EC}(t) = \lambda^{(\gamma)}_{EC}\,(1 + \sin 2
 \theta\,\cos(\Delta E_{m'm''}t)).
\end{eqnarray}
For the calculation of the transition rate $\lambda^{(\gamma)}_{EC}$
of atom ${^{180}}{\rm Re}$ we have to take into account not only
K--shell electron captures but also L--, M--, N-- and so on shell
electron captures. According to \cite{STW}, the transition rate
$\lambda^{(\gamma)}_{EC}$ is proportional to the wave function
$|\psi_{ns}(0)|^2 = Z^3\alpha^3m^3_e/n^3$, summed over the {\it
principal} quantum number $n$. For atom ${^{180}}{\rm Re}$ the
contribution of the L--, M--, N-- and so on shell electron captures
relative to the K--shell electron capture makes up of about $19\,\%$,
since the upper shell corresponds to $n = 6$ \cite{PDG08}. However,
the contribution of the electron captures from the higher shells does
not influence the frequency of the time modulation.

Thus, the explanation of the ``GSI oscillations'', based on the
hypothesis of the existence of two closely spaced ground
mass--eigenstates of the mother nuclei, predicts the time modulation
of the $EC$--decay rates of the ${^{180}}{\rm Re}_{I^{\pi} = 1^-} \to
{^{180}}{\rm W}_{I^{\pi} = 2^-} + \nu_e \to {^{180}}{\rm W}_{I^{\pi} =
2^+} + \gamma + \nu_e$ and ${^{142}}{\rm Pm}_{I^{\pi} = 1^+} \to
{^{142}}{\rm Nd^*}_{I^{\pi} = 0^+} + \nu_e \to {^{142}}{\rm
Nd}_{I^{\pi} = 0^+} + \gamma + \nu_e$ decays with a period of about
$T_{EC} \simeq 5\,{\rm s}$. The time spectra for ${^{180}}{\rm Re}$
and ${^{142}}{\rm Pm}$, obtained after a single irradiation by setting
a narrow gate on the $903\,{\rm keV}$ gamma line \cite{GSI6} and for
${^{142}}{\rm Nd}$ $K_{\alpha}$ $X$--rays, are shown in Fig.\,1 and
Fig.\,2, respectively.  The time axes in Fig.\,1 and Fig.\,2 are
binned with $1\,{\rm s/\rm bin}$ and $0.5\,{\rm s/\rm bin}$,
respectively.  During $1\,{\rm s/\rm bin}$ and as well as $0.5\,{\rm
s/\rm bin}$ the time modulation of the $EC$--decay rates of atoms
${^{180}}{\rm Re}$ and ${^{142}}{\rm Pm}$ with periods $T_{EC} \simeq
5\,{\rm s}$ should be measured well. However, such a time modulation
is not shown on the experimental time spectra and the prediction of
such a time modulation disagrees with the experimental data
\cite{GSI6,PV08}.

\subsubsection*{$EC$--decay of mother  atoms ${^{180}}{\rm
Re}$ and ${^{142}}{\rm Pm}$ in a theory of weak interactions with
massive neutrinos}

In a theory of weak interactions with massive neutrinos a time
modulation of the $EC$--decay rates of atoms ${^{180}}{\rm Re}$ and
${^{142}}{\rm Pm}$ can appear only due to a coherence of the processes
of the emission of massive neutrinos $m \to d_- + \nu_1 \to d_+ +
\gamma + \nu_1$ and $m \to d_- + \nu_2 \to d_+ + \gamma + \nu_2$.
This is unlike the approach, based on the existence of two closely
spaced mass--eigenstates of the mother nuclei $m$, where a coherence
in the initial state Eq.(\ref{label3}) cannot be destroyed by the
subsequent transitions $m \to d_- + \nu_e \to d_+ + \gamma + \nu_e$.

A coherence of the processes of the emission of massive neutrinos $m
\to d_- + \nu_1 \to d_+ + \gamma + \nu_1$ and $m \to d_- + \nu_2 \to
d_+ + \gamma + \nu_2$ can be destroyed by the broadening of neutrino
mass--eigenstates energies caused by (i) a very short lifetime of the
K--hole, created in the electron capture process, (ii) short lifetimes
of the excited nuclear states, (iii) the excited phonon spectra in the
solid target and so on. So there exist enough physical reasons to
argue that in a theory of weak interactions of massive neutrinos the
$EC$--decay rates of atoms ${^{180}}{\rm Re}$ and ${^{142}}{\rm Pm}$,
measured in \cite{GSI6,PV08}, should not show any time modulation.

Nevertheless, we would like to show that even if a coherence of the
processes of the emission of massive neutrinos is not destroyed, the
$EC$--decay rates of  atoms ${^{180}}{\rm Re}$ and
${^{142}}{\rm Pm}$, calculated in such an approach, should not show an
observable time modulation.

In the approach, assuming the interference of massive neutrinos for
the explanation of the ``GSI oscillations'', the integrand of the
$EC$--decay rate Eq.(\ref{label7}) of the $m \to d_- + \nu_e \to d_+ +
\gamma + \nu_e$ decay can be obtained in analogy with that given by
Eq.(\ref{label8}). The interference term takes the form $ 2
U_{e2}U_{e1}\cos (\omega_{EC}t) = \sin 2\theta_{12}\,\cos
(\omega_{EC}t)$, where $\theta_{12}$ is the mixing angle \cite{PDG08}
and $\omega_{EC}$ is the energy difference of neutrino
mass--eigenstates $\nu_2 $ and $\nu_1$ equal to $\omega_{EC} =
E_2(\vec{k} + \vec{k}_+) - E_1(\vec{k} + \vec{k}_+)$.  The massive
neutrinos have equal 3--momenta due to momentum conservation,
described by the $\delta$--function $\delta^{(3)}(\vec{k} + \vec{k}_+
+ \vec{k}_j)$, where $\vec{k}_j$ is a 3--momentum of neutrino
mass--eigenstate $\nu_j$.  Since 3--momenta of massive neutrinos are
equal, the energy difference $\omega_{EC} = E_2(\vec{k} + \vec{k}_+) -
E_1(\vec{k} + \vec{k}_+)$ is inversely proportional to the $Q$--value
of the $EC$--decay
\begin{eqnarray}\label{label10}
\hspace{-0.3in}&&\omega_{EC} = E_2(\vec{k} + \vec{k}_+) - E_1(\vec{k}
+ \vec{k}_+) =\nonumber\\\hspace{-0.3in} &&= \frac{m^2_2 -
m^2_1}{E_2(\vec{k} + \vec{k}_+) + E_1(\vec{k} + \vec{k}_+)} \simeq
\frac{\Delta m^2_{21}}{2 Q_{EC}},
\end{eqnarray}
where $Q_{EC} = 3800\,{\rm keV}$ and $Q_{EC} = 4870\,{\rm keV}$ are
the $Q$--values of the $EC$--decays of atoms ${^{180}}{\rm Re}$ and
${^{142}}{\rm Pm}$ \cite{ND96}, respectively, and $E_2(\vec{k} +
\vec{k}_+) \simeq E_1(\vec{k} + \vec{k}_+) \simeq Q_{EC}$
\cite{Ivanov4,Ivanov1}.

Thus, the rates of the ${^{180}}{\rm Re}_{I^{\pi} = 1^-} \to
{^{180}}{\rm W}_{I^{\pi} = 2^-} + \nu_e \to {^{180}}{\rm W}_{I^{\pi} =
2^+} + \gamma + \nu_e$ and ${^{142}}{\rm Pm}_{I^{\pi} = 1^+} \to
{^{142}}{\rm Nd^*}_{I^{\pi} = 0^+} + \nu_e \to {^{142}}{\rm
Nd}_{I^{\pi} = 0^+} + \gamma + \nu_e$ decays in the theory of weak
interactions with massive neutrinos read
\begin{eqnarray}\label{label11}
\lambda^{(\gamma)}_{EC}(t) = \lambda^{(\gamma)}_{EC} (1 + \sin
2\theta_{12}\,\cos(\omega_{EC}t)),
\end{eqnarray}
where the frequencies are equal to
\begin{eqnarray}\label{label12}
\omega_{EC} = \left\{\begin{array}{r@{\;,\;}l}
4.38\times 10^4\,{\rm s^{-1}} & {^{180}}{\rm Re}\\
3.42\times 10^4\,{\rm s^{-1}} & {^{142}}{\rm Pm},
\end{array}\right.
\end{eqnarray}
calculated for
$\Delta m^2_{21} = 2.19\times 10^{-4}\,{\rm eV^2}$ \cite{Ivanov3}.
The periods of time modulation are
\begin{eqnarray}\label{label13}
T_{EC} = \frac{2\pi}{\omega_{EC}}= \left\{\begin{array}{r@{\;,\;}l}
1.43\times 10^{-4}\,{\rm s} & {^{180}}{\rm Re}\\
1.84\times 10^{-4}\,{\rm s} & {^{142}}{\rm Pm},
\end{array}\right.
\end{eqnarray}
During $1\,{\rm s/\rm bin}$ and $0.5 \,{\rm s/\rm bin}$ the periodic
terms of the $EC$--decay rates of atoms ${^{180}}{\rm Re}$ and
${^{142}}{\rm Pm}$, calculated in a theory of weak interactions with
neutrino mass--eigenstates, make of about $7000$ and $2700$
oscillations, respectively, and, of course, cannot be observed by the
reported experiments. This agrees well with the experimental time
spectra in Fig.\,1 and Fig.\,2, measured in \cite{GSI6,PV08}.

\subsection*{Conclusive discussion}

We have shown that following the hypothesis of the existence of two
closely spaced ground mass--eigenstates of the mother nuclei of atoms
${^{180}}{\rm Re}$ and ${^{142}}{\rm Pm}$ one gets the $EC$--decay
rates of atoms ${^{180}}{\rm Re}$ and ${^{142}}{\rm Pm}$, modulated
with periods $T_{EC}\simeq 5\,{\rm s}$. This disagrees with the
experimental data, obtained in \cite{GSI6,PV08}.

The absence of time modulation of the $EC$--decay rates of atoms
${^{180}}{\rm Re}$ and ${^{142}}{\rm Pm}$, calculated in a theory of
weak interactions with neutrino mass--eigenstates, can be related to a
violation of a coherence of the processes of the emission of massive
neutrinos. Such a decoherence can have some physical reasons caused by
short lifetimes of the K--holes, produced by the captured electrons,
excited nuclei, excited phonon spectra of solid targets and so on
\cite{GSI7}.

Nevertheless, we have shown that even if a coherence of the processes
of the emission of massive neutrinos is not destroyed, a time
modulation of the $EC$--decay rates of atoms ${^{180}}{\rm Re}$ and
${^{142}}{\rm Pm}$ appears with periods of order of $T_{EC} \sim
10^{-4}\,{\rm s}$, which cannot be observed at the present
experiments. This agrees well with the experimental data on the time
spectra of the $EC$--decays of ${^{180}}{\rm Re}$ and ${^{142}}{\rm
Pm}$ atoms \cite{GSI6,PV08}.

Since the experimental data on the time modulation of the number of
the $EC$--decay rates of ${^{180}}{\rm Re}$ have declared as the test
for the theoretical approaches for the explanation of the ``GSI
oscillations'', measured in \cite{GSI2}--\cite{GSI5}, the results,
obtained above, allow us to argue that the time modulation of the
$EC$--decay rates of the H--like heavy ions ${^{142}}{\rm Pm}^{60+}$,
${^{140}}{\rm Pr}^{58+}$ and ${^{122}}{\rm I}^{52+}$ and the
$\beta^+$--decay rates of the H--like ${^{142}}{\rm Pm}^{60+}$ can be
explained only due to the interference of neutrino mass--eigenstates
\cite{Ivanov4} (see also \cite{Ivanov2}).

\end{document}